\documentstyle[12pt,psfig]{l-aa}
\def \src {XB\thinspace1323$-$619}

\def \sax {BeppoSAX}
\def \degmark{^\circ}
\def \nh {N${\rm _H}$}
\def \ergsec{\hbox{erg s$^{-1}$}}
\def \hcm {\hbox {\ifmmode $ atoms cm$^{-2}\else atoms cm$^{-2}$\fi}}
\def \arcmin {\hbox{$^\prime$}}

\def\approxgt{\mathrel{\hbox{\rlap{\lower.55ex \hbox {$\sim$}}
        \kern-.3em \raise.4ex \hbox{$>$}}}}
\def\approxlt{\mathrel{\hbox{\rlap{\lower.55ex \hbox {$\sim$}}
        \kern-.3em \raise.4ex \hbox{$<$}}}}
\input psfig.sty
\begin{document}

   \thesaurus{6(13.25.5; 
               08.09.2: XB\thinspace 1323-619;  
               08.14.1;  
               08.02.1;  
               02.01.2)} 
\title{An X-ray study of the dipping low mass X-ray binary 
       \src}

\author{M. Ba\l uci\'nska-Church\inst{1}
 \and M.J. Church\inst{1}
 \and T. Oosterbroek\inst{2}
 \and A. Segreto\inst{3}
 \and R. Morley\inst{1}
 \and A.N. Parmar\inst{2}
}
   \offprints{M. Ba\l uci\'nska-Church (mbc@star.sr.bham.ac.uk)}
   \institute{School of Physics and Astronomy, University of
Birmingham,
              Birmingham, B15 2TT, UK
   \and
              Astrophysics Division, Space Science Department of ESA,
ESTEC,
              Postbus 299, NL-2200 AG Noordwijk, The Netherlands
   \and
              Instituto IFCAI, via La~Malfa 153, I-90146 Palermo, Italy
}

\date{Received 4 March 1999; accepted 25 June 1999}

\maketitle

\markboth{Study of \src\ }{Study of \src\ }

\begin{abstract}

During a \sax\ observation of the low-mass X-ray binary dip source \src\ 
a total of 10 type I X-ray bursts and parts of 12 intensity dips were 
observed. During non-bursting, non-dipping intervals, the 1.0--150~keV \sax\ 
spectrum can be modelled by a cutoff power-law with a photon index 
of $\rm {1.48 \pm 0.01}$, a cutoff energy of $\rm
{44.1^{+5.1}_{-4.4}}$ keV together with 
a blackbody with kT of $\rm {1.77 \pm 0.25}$ keV contributing
$\sim$15\% of the 2 -- 10 keV flux. Absorption equivalent to
$\rm {3.88 \pm 0.16\times 10^{22}}$~atom~cm$^{-2}$ is required. 
The dips repeat with a period of $2.938 \pm 0.020$~hr and span 40\%
of the orbital cycle. During dips the maximum reduction in 2--10~keV
intensity is $\sim$65\%. The spectral changes during dips are complex and cannot be
modelled by a simple absorber because of the clear presence of part of
the non-dip spectrum which is not absorbed. Spectral evolution in
dipping can be well modelled by progressive covering of the 
cutoff power-law component which must be extended, plus rapid absorption 
of the point-source blackbody. One of the bursts 
is double and 4 of the bursts occurred during dipping intervals. These bursts 
have 2--10~keV peak count rates reduced by only 22\% on average from those 
occurring outside the dips, and are not heavily absorbed. 
One explanation for this lack of absorption is that 
the bursts temporarily ionize the absorbing material responsible for the dips.

\end{abstract}


\section{Introduction}

A significant fraction of low-mass X-ray binaries (LMXB) exhibit periodic dips 
in their X-ray intensity. The dips recur at the orbital period of the system 
and are believed to be caused by periodic obscuration of a central X-ray
source by structure located in the outer regions of an
accretion disk (White \&
Swank 1982). The depth, duration and spectral properties of the dips 
vary from source to source and from cycle to cycle (e.g. Parmar \& White
1988). Determination of the spectral evolution in dipping allows not only
the structure and dynamics of the outer regions of the accretion
disk and the properties of the absorbing material, but also the 
geometry and nature of the emission regions, to be investigated
(see Church et al. 1997, 1998a,b). 
In the LMXB dip sources, the spectral changes during dips are complex
and can not be described by simple absorption of one-component emission
but can be described by assuming two emission components: one
point-like blackbody from the surface of the neutron star, and the
other extended Comptonized emission from an Accretion
Disk Corona (ADC) (Church \& Ba\l uci\'nska-Church 1993, 1995). 
In some sources, dipping is well explained in terms of a relatively
small absorber which covers the point-like emission but does not
absorb the extended emission to a large extent. In another group of 
dip sources, the dip spectra indicate the presence of excesses at
$\approxlt$4~keV, i.e. clear evidence that part of the spectrum is not
absorbed (Parmar et al. 1986; Courvoisier et al. 1986; Smale et al.
1992; Church et al. 1997). \src\ belongs to this group (Parmar et al.
1989). The dip spectra of these sources have often been fitted 
by a technique in which the non-dip spectral form is split into two parts
each having the form of non-dip emission, but only one of
which suffers extra absorption, while the other has a normalization
which decreases markedly in dipping, often by a factor of 10.
This may be called `absorbed + unabsorbed' modelling.
Explanations for the apparent change in normalization have been based on 
electron scattering, partial covering by the absorber and fast variations of column density
(Parmar et al. 1986); however, it has been difficult to explain the
large changes convincingly. For example, a large degree of 
electron scattering is unlikely
in the outer accretion disk where the ionization state is low as
demonstrated by the observed occurrence of photoelectric absorption in dipping.
An alternative approach based on the
two-component model (above) for the dip sources has been suggested 
(Church et al. 1997). In this group of sources, the absorber is extended and
often of larger angular extent than all emission regions. Thus in
dipping, the point-like blackbody region is covered immediately
whereas the extended Comptonized emission region is progressively
covered by the absorber. This approach can explain the complex
behaviour in several sources (Church et al. 1997, 1998a,b), and
provides a simple explanation in which the unabsorbed part of the dip spectrum  
originates from the uncovered emission as the extended absorber moves
across the extended emission region.

\src\ is a faint ($\sim$3 mCrab), poorly studied LMXB that
exhibits 2.93~hour periodic intensity dips
and X-ray bursts. The source was first detected by {\it Uhuru}
and {\it Ariel V} (Forman et al. 1978; Warwick et al. 1981) and
the dips, bursts and a soft excess during dips were discovered using EXOSAT
(van der Klis et al. 1985; Parmar et al. 1989, hereafter P89).
Recently, $\sim$1 Hz quasi-periodic oscillations have been discovered
in the persistent emission, the dips and the bursts from \src\ (Jonker et al.
1999). 
The EXOSAT Medium Energy 1--10~keV non-dip spectrum could be modelled
by either a power-law with a photon index of $\rm {1.53 \pm 0.07}$ and
absorption, \nh\, equivalent to $\rm {(4.0 \pm 0.3)\times 10^{22}}$ atom
$\rm {cm^{-2}}$, or a thermal bremsstrahlung with a temperature of $\rm
{26^{+14}_{-6}}$ keV (P89).
During the dips, which typically last for 40\% of the
orbital cycle, the X-ray intensity varies
irregularly with a minimum of $\sim$50\% of the average value outside of
the dips. During the EXOSAT observation the bursts repeated every
5.30--5.43~hr. 
\src\ lies close to the galactic plane (l$_{II} = 307.0\degmark$,
b$_{II} = 0.4\degmark$ and the average column density in this direction
is $\rm {\sim 1.4 \times 10^{22}}$~atom~cm$^{-2}$ (using Stark et al.
1995).
The region of sky containing \src\ is complex with ten point sources
detected in previous EXOSAT Channel Multiplier Array (CMA)
and {\it Einstein} imaging observations
(see Fig.~4 of P89). On the basis of an extreme
hardness ratio, P89 proposed that source D, detected during
{\it Einstein} Imaging Proportional Counter (IPC) and High Resolution
Imager observations (HRI), is probably \src. The count rates in these
observations are too low to detect dipping clearly.
The identification of source D with \src\ was recently
confirmed when Smale (1995) discovered a faint ($K' = 17$)
variable IR object within the uncertainty region for source D
during a short (1.5~hr) observation using the Infrared Imaging
Spectrometer at the Anglo-Australian Telescope.

We present a detailed study of the \src\ X-ray spectrum and its
evolution during dips using results obtained with \sax. We compare
the properties of bursts that occurred during dipping with those outside 
dipping intervals. Spectral fitting results from the previously unpublished 
1994 ASCA observation of \src\ are presented and compared with the \sax\ 
results. The serendipitous detection of a pulsing source within the
\src\ field using \sax\ is reported in Angelini et al. (1998).

\section{Observations}
\label{sec:observations}

\subsection{\sax}
\label{subsect:sax}

Data from the Low-Energy Concentrator Spectrometer (LECS;
0.1--10~keV; Parmar et al. 1997), Medium-Energy Concentrator
Spectrometer (MECS; 1.3--10~keV; Boella et al. 1997),
High Pressure Gas Scintillation Proportional Counter
(HPGSPC; 5--120~keV; Manzo et al. 1997) and the Phoswich
Detection System (PDS; 15--300~keV; Frontera et al. 1997) on-board \sax\
are presented. All these instruments are coaligned and collectively referred
to as the Narrow Field Instruments, or NFI.
The MECS consists of three identical grazing incidence
telescopes with imaging gas scintillation proportional counters in
their focal planes, however prior to the observation of \src\ one of
the detectors had failed. The LECS uses an identical concentrator system as
the MECS, but utilizes an ultra-thin entrance window and
a driftless configuration to extend the low-energy response to
0.1~keV. The non-imaging HPGSPC consists of a single unit with a collimator
that is alternatively rocked on- and off-source to monitor the
background spectrum. The non-imaging
PDS consists of four independent units arranged in pairs each having a
separate collimator. Each collimator can be alternatively
rocked on- and off-source to monitor the background.

\begin{figure*}
  \centerline{\psfig{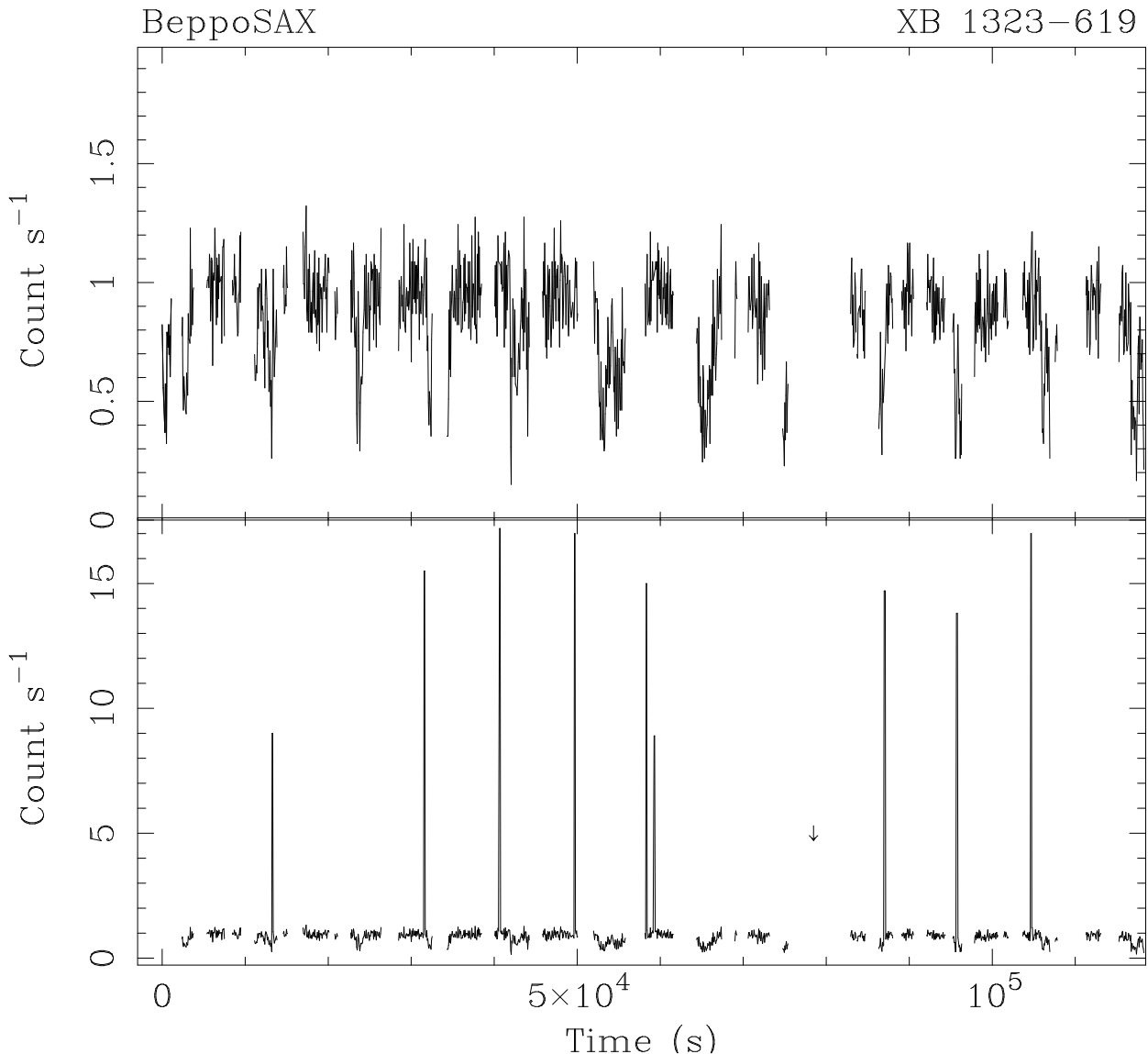}}
    \caption[]{MECS 2--10 keV lightcurves of \src\ with a binning
                 time of 64~s. In the upper panel the bursts are removed
                              in order to show the dips clearly. The
position of the
             burst observed by the LECS but not by the MECS is
                          indicated by an arrow}
                            \label{fig:lightcurve}
                            \end{figure*}

The region of sky containing \src\ was observed by \sax\
between 1997 August 22 17:06 and August 24 02:02~UTC.
Good data were selected from intervals when the elevation angle
above the Earth's limb was $>$$4^{\circ}$ and when the instrument
configurations were nominal, using the SAXDAS 1.3.0 data analysis package.
The standard collimator dwell time of 96~s for each on- and
off-source position was used, together with rocking angles of 180\arcmin\ 
and 210\arcmin\ for the HPGSPC and PDS, respectively. The exposures 
in the LECS, MECS, HPGSPC, and PDS instruments are 15~ks, 70~ks,
28~ks, and 28~ks, respectively. LECS and MECS data were
extracted centered on the position of \src\ using radii of 8\arcmin\ and
4\arcmin, respectively. Background subtraction in the imaging instruments
was performed using standard files, but is not critical for such a
bright source. Background subtraction in the non-imaging instruments was 
carried out using data from the offset intervals.

\subsection{ASCA}

The ASCA instrumentation consists of two Solid State Imaging
Spectrometers SIS0 and SIS1 (0.6--10~keV), and two Gas Imaging
Spectrometers GIS2 and GIS3 (0.8--10~keV, Tanaka et al. 1994). The energy resolution
of the GIS is similar to that of the \sax\ LECS and MECS in the overlapping
energy range, while that of the SIS is a factor of a few better, except at
the lowest energies.

ASCA observed the \src\ region on 1994 August 04 between 12:28 and
20:10~UTC with a GIS exposure of 22~ks. GIS data were screened to remove regions of
high particle background, to restrict elevation above the rim of the
Earth to $>$$5^{\circ}$, particle rigidity to more than 6 GeV~c$^{-1}$, the
radiation belt parameter to less than 200 count~s$^{-1}$, and angular deviation to less
than $0.014^{\circ}$. The calibration source and outer ring were removed from
the image, and rise-time rejection applied. Source and background data
were selected from diametrically opposite 6\arcmin\ radius regions.
Two bursts and parts of two dips were observed.

\section{Results}

\subsection {The X-ray lightcurve}

Figure~\ref{fig:lightcurve} shows the background-subtracted 2--10~keV MECS lightcurves
of \src\ with a binning of 64~s. The scaling is chosen to show the dipping clearly 
in the upper panel and the bursting clearly in the lower panel. 
Parts of 12 dips were 
seen as irregular reductions in intensity as well as 10 X-ray bursts. One of these 
is a double burst, and one burst was observed by the LECS, but not by the MECS which 
was switched-off for one satellite orbit during the observation for technical reasons.
With the binning used, the apparent heights of the bursts (lasting $\sim$100~s) are
substantially reduced from their true heights obtained from a light
curve with much shorter binning (less than 4 s). Since we wish to show clearly the
differences between the bursts that occur in non-dip emission and those coinciding with
dips (Sect. 4), we have corrected the burst heights in Fig. 1 to their values
with binning less than 4 s.
Bursts and dips are not strongly evident in the HPGSPC and PDS lightcurves due to the 
poorer signal to noise ratio of these instruments and the energy dependence of these 
features.

It can be seen from Fig.~1 that dipping is not 100\% deep in the band 2--10~keV, 
and so the source differs from XB\thinspace 1916-053, for example 
(Church et al. 1998b), in which dipping is usually 100\%. To investigate
this, light curves were also extracted in various energy bands within the
range of the MECS. The results showed dipping to be 
100\% deep (at its deepest point) in the band
below 2.3~keV, decreasing systematically with
increasing energy, i.e. 67\% in the band 3--6~keV and 57\% in the band 6--10~keV.
However, the count rate was low, particularly in the lowest band and the
Poisson errors large. If dipping is 100\% deep at the lowest energies,
the consequence is that the overall size of the absorber must be larger
than the largest emission region. This is further discussed in Sect. 5.

\begin{figure}
  \centerline{\psfig{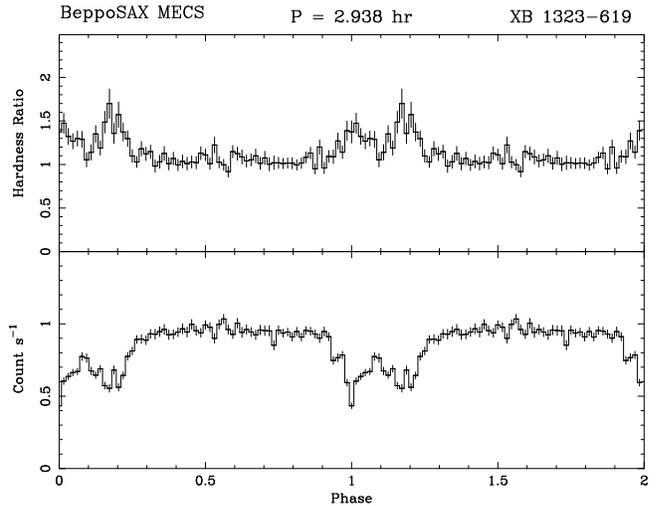}}
  \caption[]{The \src\ lightcurve and hardness ratio folded on the
             dip period  and displayed over repeated cycles for
             clarity.
             Intervals corresponding to X-ray bursts are
             excluded. Dipping activity is clearly seen as increases
             in hardness ratio}
  \label{fig:folded}
\end{figure}

Figure~\ref{fig:folded} shows the 2.0--10~keV background
subtracted MECS lightcurve and hardness ratio folded on the
best-fit dip period given below. 
Intervals corresponding to X-ray bursts have
been excluded. The hardness ratio is defined as the count rate in the energy 
range 4.5--10.0~keV divided by that in the range 
1.8--4.5~keV. Dipping activity is clearly seen as increases in hardness
ratio consistent with the energy-dependence of dipping
and is evident between phases 0.9 to 1.3 (where
phase 0.0 is defined as the epoch of deepest dipping).

The dip period was determined Using MECS data from
which the bursts were excluded,  and found to be 2.938 $\pm$ 0.020~hr.
The period and 1$\sigma$ uncertainty were derived by
estimating the times corresponding to the center of each dip from
template fits and fitting the resulting arrival times to a linear
relation with cycle number. The folded templates were updated as more
refined values of the period were obtained until the fitting converged.
The minimum in the folded light curve occurs at MJD\thinspace
50682.719 $\pm$ 0.0045. The period is consistent with that derived from the
1985 February EXOSAT observation of 2.932 $\pm$ 0.005~hr (P89). ASCA observed
only two dips, one of which was broken by data gaps, and the dip period
cannot be reliably obtained.

\subsection {X-ray spectra}

The spectrum of non-dip emission was investigated by simultaneously
fitting data from all the \sax\ NFI. Time intervals corresponding to
bursts were first removed, and then data selected with phases between
0.3 and 0.9 using the ephemeris given in Sect. 3.1. This gives
background-subtracted count rates of 0.41, 1.12, 2.59 and 2.16 count
s$^{-1}$ for the LECS, MECS, HPGSPC and PDS, respectively.
Spectral evolution in dipping was investigated using MECS data only
since the dips are not strongly seen at higher energies and there are too
few counts in the LECS to contribute meaningfully. 
After initial trials, MECS dip spectra were extracted with count rates
between 0.3--0.6 count s$^{-1}$, 0.6--0.75 count s$^{-1}$ and 0.75--0.85 count
s$^{-1}$ after first binning the data in 128s intervals.  Given the
count rate of the source, a larger number of dip spectra would not be
sensible.
The LECS and MECS spectra were rebinned to oversample the full width
at half maximum of the energy resolution by a factor of 3, and
additionally  LECS  data were rebinned to a minimum of 20 counts per bin 
and MECS data to 60 counts per bin to allow use of the $\chi^2$ statistic.
LECS data were only used between 1--5 keV and MECS data between
1.8--10.0 keV where the instrument responses are well determined. 
The HPGSPC and PDS
data were rebinned using standard binnings in the bands 7--20~keV and
13--150~keV, respectively. In the following, the photoelectric absorption
cross sections of Morrison \& McCammon (1983) were used incorporating the
Solar abundances of Anders \& Grevesse (1989).

\begin{table*}
\caption[]{Spectral fitting results obtained by simultaneously fitting
the \sax\
NFI non-dip spectrum and 3 dip spectra using various models (see text).
\nh\ is in units of $\rm {10^{22}}$ atom $\rm {cm^{-2}}$.
90\% confidence limits are given}
\begin{flushleft}
\begin{tabular}{llllll}
\hline\noalign{\smallskip}
Model & \hfil N$_{\rm {H}}$ \hfil & kT (keV) &\hfil $\alpha$ \hfil
& $\rm{E_{co}}$ (keV) & $\chi^2$/dof \\
\noalign{\smallskip\hrule\smallskip}
Power-law & $4.71\pm 0.18$ & \dots & $1.81\pm 0.01$ & \dots &
1110/397  \\
Bremsstrahlung & $3.52\pm 0.17$ & $23.4\pm 1.88$ & \dots & \dots &
995/396\\
Cutoff power-law (simple absorber)& $3.98\pm 0.18$ & \dots &$ 1.43\pm
0.01$ &
$40.9 ^{+4.2}_{-3.7}$ & 665/396\\
Cutoff power-law + cutoff power-law &$4.05\pm 0.16$ & \dots & $1.43\pm
0.01$&$40.5^{+4.1}_{-3.6}$&392/390 \\
Cutoff power-law (prog. covering)& $4.26\pm 0.16$ & \dots &$1.49\pm 0.01$
&$44.0^{+4.9}_{-4.2}$ &453/393\\
Cutoff power-law + blackbody (prog. covering)& $3.88\pm 0.16$ & $1.77\pm
0.25$
& $ 1.48\pm 0.01$ & $44.1^{+5.1}_{-4.4}$ & 401/387 \\
\noalign{\smallskip}
\hline
\end{tabular}
\end{flushleft}
\label{tab:spec_paras}
\end{table*}

\begin{figure*}
  \centerline{\psfig{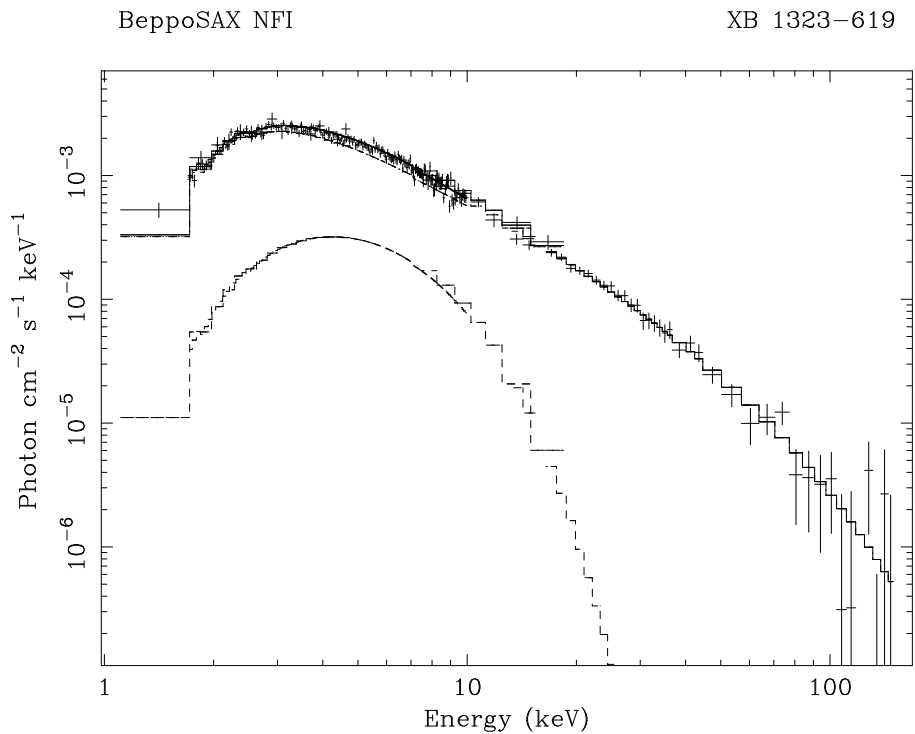}}
        \caption[]{The non-dip NFI \src\ spectrum fitted with the
absorbed
                 cutoff power-law and blackbody model discussed in the
text.
                 The data have been rebinned to aid clarity. The total
model
                 and the contribution of the blackbody component are
shown
                 separately}
                 \label{fig:broadband}
                 \end{figure*}
                 
In order to understand the spectrum of a dip source, the spectral
model must give acceptable fits to both the non-dip
and dip spectra. Since dips are almost certainly due to obscuration
by intervening material, a satisfactory model should fit the
dip spectra without requiring any of the parameters that 
characterize the source emission such as the temperature or power-law
index to be changed. Normally, the procedure followed consists of
obtaining these parameters from fitting the non-dip spectrum. However,
in the case of weak sources, the non-dip spectrum
alone may not constrain parameters sufficiently. This is the case here,
and instead, the NFI non-dip spectra and the 3 dip spectra were fitted
simultaneously. This approach proved successful in discriminating between
various models, and providing a good fit to all spectra with our best model. 

Initially, simple models were tried, including absorbed power-law,
thermal bremsstrahlung and cutoff power-law: ${\rm
E^{-\alpha}\exp-(E/E_{co}}$) models. 
Factors were included in the spectral fitting to allow for normalization 
uncertainties between the instruments;  values of these factors were
similar to those found for other sources. The power-law model 
gave an unacceptable fit with a $\chi ^2$  of 1110/397, and strong discrepancies
between model and data above 40~keV where there is down-curving of the
spectrum.
The bremsstrahlung model similarly did not provide an acceptable fit
with a $\chi^2$ of 995/396.
Although the fit to the non-dip data alone was only marginally unacceptable, 
there were large discrepancies between data and model for deeper dip data. 
The cutoff power-law was also unacceptable with a $\chi^2$ of
665/396. In all cases, there is an unabsorbed component in the 
spectra at low energies (or soft excess), most obvious in the deeper dip spectra. Thus, 
there was little change in the low energy cutoff of the spectrum 
in dipping compared with the non-dip spectrum.

\begin{figure*}[!t]
  \centerline{\psfig{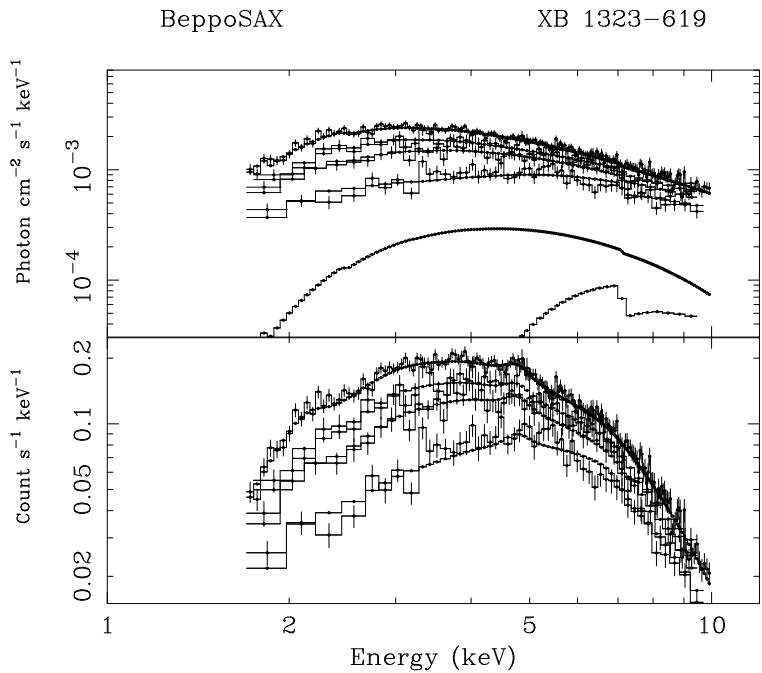}}
      \caption[]{Progressive covering fits to the MECS non-dip and 3 dip
          spectra
                      given in Table 2. The lower panel shows the
                                          spectra and total model
folded through the
instrument response.
        The upper panel shows the observed counts spectra with total
        model
                and the blackbody components (lower curves), omitting
the
                cutoff power-law for clarity. Only the blackbody
                contributions to
                        the non-dip and shallow dip spectrum are
shown, as the
        contributions
                        in the other cases are too small to be included}
                                          \label{fig:dips}

\end{figure*}

Next, the model frequently applied to the dip sources with an
unabsorbed component at low energies was tried (e.g. Parmar et al. 1986). 
The non-dip model consisting of a cutoff power-law was split for 
the dip spectra into two cutoff power-laws having identical 
spectral parameters except for the column densities and normalizations.
In the one component, the column density was allowed to vary, in the
other, the column density was frozen at the non-dip value; both
normalizations were free. For the non-dip spectrum, this model reduces to 
a single component. This model fitted the spectra well with an overall 
$\chi^2$ for the fit of 392/390. The absorbed component has an increasing
\nh\ in dipping with approximately constant normalization; however
for the unabsorbed component there is a large change in normalization
which decreases by a factor of $\sim$2 in the deepest dip spectrum of 
Fig.~4 (intensity band 0.3 - 0.6 count s$^{-1}$). We also sub-divided 
this band making a spectrum for 0.3 - 0.4 count s$^{-1}$ for which the 
normalization change was 2.5. We are not able to explain such a decrease in
normalization: for example, in Fig.~5, there is no evidence that
extensive electron scattering of X-rays takes place in the absorber which
would be seen as an energy-independent shift between the deep dip and
non-dip spectra shown. 

However, for the first time, \sax\ offers the ability to test this model 
more rigorously since the normalization decrease, if real, implies
that the PDS intensity (at energies where absorption plays no part) would have 
to decrease substantially between non-dip and deep dipping, which it 
clearly does not. Fitting the deep dip model obtained above to the
PDS deep dip spectrum has a $\chi^2$ of 27.4/7 for the data between 20
and 50 keV. Thus, this model is not able to fit the PDS data. The present 
work on \src\, and previous work on other sources with an unabsorbed 
part of the spectrum 
(Church et al. 1997, 1998a,b) where the normalization decrease is typically 
a factor of 10, provide evidence that the change in normalization 
is unreal, and is due to the application of an inappropriate model.
Thus this fitting is shown in Table 1 only for completeness, as a cutoff 
power-law + cutoff power-law model. 

\begin{table*}
\caption[]{Best fits to the dip spectra.
$\rm {N_H}$ is in units of $\rm {10^{22}}$ atom $\rm {cm^{-2}}$
for both the spectral components. 90\% confidence limits are given}
\begin{flushleft}
\begin{tabular}{llllll}
\noalign{\hrule\smallskip}
Spectrum &MECS ct &$\rm {N_H^{BB}}$ &$\rm {N_H^{CPL}}$ &\hfil f \hfil&
$\chi^2$/dof\\
&rate (s$^{-1}$) \\
\noalign{\smallskip\hrule\smallskip}
Non-dip &\dots      &3.88$\pm 0.16$       &3.88$\pm 0.16$  &0.0
&150/163 \\
Shallow dip &0.75--0.85  &$63^{+110}_{-28}$         &7.0$\pm 1.1$
&0.376$\pm 0.085$ &41/48 \\
Medium dip &0.60-0.75   &$\rm {2.2\times 10^4}$  &13.8$\pm 1.8$
&0.423$\pm 0.036$ & 46/48 \\
Deep dip &0.30--0.60  &$\rm {6.1\times 10^4}$  &26.7$\pm 2.1$
&0.665$\pm 0.019$ & 75/51 \\
\noalign{\smallskip}
\hline
\end{tabular}
\end{flushleft}
\label{tab:dips}
\end{table*}

We next tried the progressive covering model, which  consists of a point-like 
blackbody and an extended Comptonization term, with progressive covering of
the extended component in dipping (e.g. Church et al. 1998b). The model flux 
may be written:
\[
{\rm e^{-\sigma_{MM}N_H} \;( I_{BB}  e^{-\sigma_{MM}
N_H^{BB}} \rm +\;  I_{CPL}\,(f\, e^{-\sigma_{MM}N_H^{CPL}}+\; (1 - f))}
\]
Here $\rm {I_{BB}}$ and $\rm {I_{CPL}}$ are the normalizations of the
blackbody and cutoff power-law components,
${\rm N_H ^{BB}}$ and ${\rm N_H^{CPL}}$ are the column densities for each
component during dipping (additional to the non-dip $\rm
{N_H}$), and {\it f} is the covering fraction.
This model gave an acceptable $\chi^2$ of 401/387. 
The cutoff
power-law parameters are provided essentially by the deep dip spectrum
in which the blackbody is totally absorbed, whereas the blackbody
parameters are provided by the non-dip spectrum.

Removing the blackbody term and re-fitting simultaneously gave a 
significantly worse $\chi^2$ of 
453/393. An F-test shows that the additional component is 
significant with $>>$99.9\% confidence. Moreover when $\chi^2$ is obtained for 
the spectra fitted individually, using the best simultaneous solution, the fits get
worse for deeper dipping with a $\chi^2$ of 100/47 in the deepest dip
spectrum. Closer examination of the non-dip and 
first dip spectra (shallow
dipping) reveals that absorption is clearly taking place in the band 5--10~keV
(see Fig. 4), indicating that a component of the spectrum 
in this energy band is rapidly removed in dipping. The 
spectral model with progressive covering of a cutoff power-law only
{\it does not allow} absorption in this band, since in shallow dipping, the
covering fraction is less than 40\%, and \nh\ for the cutoff
power-law is low (see Table 2). To produce the observed decrease in
count rate at 5~keV of 20\% for the shallow dip spectrum by progressive
covering of the cutoff power-law alone, a column density 
of more than $\rm {20\times 10^{22}}$ atom cm$^{-2}$ would be required, 
several times larger than $\rm {N_H}$ for this component.
However, when the blackbody is added,
this component being point-like is immediately covered and has high \nh\,
and so can be absorbed in the band 5--10~keV in shallow dipping.

The best-fit non-dip parameters (for the blackbody + cutoff power-law
model) are summarised in Table 1 and the fit to the 4 instruments 
shown in Fig.~3. The good fit of this
model shown in Fig.~3 confirms the existence of a high-energy spectral
break strongly supporting the Comptonized nature of the non-thermal
emission and allows parameters of the Comptonizing region to be derived.
In the 2--10~keV energy range, the blackbody contributes 14\% of the total
luminosity. The best-fit \nh\ of $\rm {(3.88 \pm 0.16)\times 10^{22}}$
atom $\rm {cm^{-2}}$ is somewhat higher than Stark et al. (1995) value of 
$\rm {\sim 1.4 \times 10^{22}}$ atom $\rm {cm^{-2}}$, suggesting the 
presence of absorber intrinsic to the source. Including a narrow Fe line 
in the non-dip fitting did not bring any improvement in fit quality, and
the 90\% confidence upper limit to the equivalent width (EW) of a narrow line 
at 6.4 keV is 48 eV. However, for the shallow and medium dip spectra,
the fits {\it were marginally improved}, with EW $<$ 100 eV and EW $<$ 160 eV in these
two cases. 

In Fig.~4, the fits of the progressive covering model (from
simultaneous fitting) to the non-dip and each of the MECS dip
spectra are shown, and the results are summarised in Table 2. During the fitting
the values of $\alpha$, kT, ${\rm E_{co}}$ and the normalizations of the two
components were chained. $\chi^2$ values were obtained by applying the
best-fit solution to each spectrum individually.
The fit to the deepest dip spectrum is not quite
so good as to the other spectra because of the reduced quality of the
spectrum. It can be seen that the
blackbody column increases rapidly, consistent with absorption of a
point-like source in the denser regions of the absorber. In 
contrast, the non-thermal component is more slowly and progressively 
covered during dipping intervals, with {\it f} rising smoothly from
zero to $\sim$0.67. The systematic increase of {\it f}
as dipping deepens is consistent with a large absorbing region moving
progressively across an extended emission component, and the unabsorbed
part of the spectrum  is simply the uncovered part of the emission.
The smaller increase of column density for the extended component
compared with the
point-like blackbody is as expected, since the blackbody will be absorbed
by denser regions in the absorber while the non-thermal component 
measures an average column density integrated across the absorber.

\subsection {ASCA spectrum}

The ASCA non-dip spectrum selected from the
GIS2 intensity band between 0.7--0.85 count~s$^{-1}$ was also
fitted using a blackbody and power-law model (the ASCA data
does not extend to high enough energies to observe the cutoff).
In
this case, $\rm {kT_{bb}}$ is $1.85 \pm ^{0.31}_{0.18}$~keV and the
photon index is $2.46 \pm ^{2.06}_{1.48}$. If a cutoff power-law is used
with $\alpha$ and $\rm {E_{co}}$ fixed at the \sax\ best-fit values,
then $\rm {kT_{bb}}$ is 1.72 $\pm \rm{^{0.38}_{0.12}}$~keV in excellent
agreement with the \sax\ results. Thus the overall spectral shape during
the ASCA observation is very similar to that observed by \sax.

\subsection {High Energy Variations}

\begin{figure}[!h]
  \centerline{\psfig{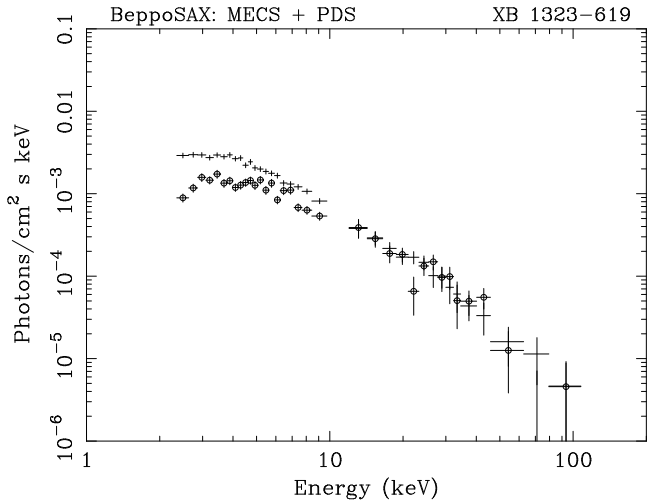}}
    \caption[]{Non-dip and deep dip (circles)  MECS and PDS spectra
                 illustrating the lack of evidence for dipping
                              $\approxgt$20~keV. The spectra have been
deconvolved using the
             best-fit progressive covering model with parameters given
                          in Tables 1 and 2}
                            \label{fig:nondeep}
                            \end{figure}
                            
Figure 4 shows that significant dipping takes place at 10~keV.
A PDS spectrum was extracted corresponding to the same time intervals used
to select the deepest dip MECS spectrum.
The mean 15--50 keV non dip PDS count rate is $\rm {1.48 \pm 0.03}$ $\rm
s^{-1}$, compared with $\rm {1.35 \pm 0.14}$ $\rm {s^{-1}}$ in deep dipping.
Fig.~5 shows the MECS and PDS non-dip and deep dip spectra, from which it
is clear that absorption continues up to $\sim$20~keV; however at higher
energies there is no evidence for any substantial energy-independent decrease of flux
in the PDS, i.e. for any large degree of electron scattering in the
absorber. A limit of $\rm {10 \pm 10}$\% made be placed on any decrease
in PDS count rate between 20 and 50 keV.

\section{X-ray bursts}

The observed bursts (and data gaps during which bursts could have 
been missed), 
are consistent with a burst recurrence timescale of between 2.40 and
2.57~hr, i.e. with regular burst occurrence as seen in the EXOSAT
observation.
This timescale is slightly less than the orbital period of the system and so
the bursts appear to ``march back'' relative to the dips during the
observation. 
During the continuous EXOSAT observation the recurrence interval of the bursts 
was significantly longer, between 5.30 and 5.43~hr (P89). 
There were also two bursts during the short ASCA observation 
separated by 3.05~hr. Although it is tempting to assume that
this difference in burst repetition rate is a consequence of a change
in the mass accretion rate, the X-ray luminosity of \src\ did not
appear to change systematically during the 3 observations.

Events for each burst were accumulated from the 
burst maxima until the intensities reached the quiescence level. 
Table 3 lists the total count and the exposure in 
each burst spectrum. The mean intensity of the bursts that occur in dips
is $\sim$70\% of those outside dips and, 
as seen in Fig.~1, the peak heights of the bursts
in dips vary between $\sim$50\% and 90\% of the non-dip burst heights.
Since there are so few counts for each burst, it is difficult to
perform individual spectral fits.

\begin{table}
\caption[]{Burst properties. Bursts 5 and 6 comprise the 
double burst and burst 7 was not observed by the MECS. 
Phase is defined using the ephemeris given in 
Sect.~3.1. Dips occur between phases 0.9--1.3}
\begin{flushleft}
\begin{tabular}{llllll}
\noalign{\hrule\smallskip} 
Burst & Time (1997)& MECS   & Exp. & Dip  & Comment  \\
No.   & Day~~~~hr:mn  & counts &  (s) & phase&         \\
\noalign{\smallskip\hrule\smallskip}
 1      & Aug 22 20:49 &   297   &  69.6  &  0.22 & Dip \\
 2      & Aug 23 01:54 &   439   &  98.3  &  0.94 & Dip \\
 3      & Aug 23 04:25 &   525   &  96.2  &  0.80 & Non-dip \\
 4      & Aug 23 06:56 &   539   &  99.3  &  0.66 & Non-dip \\
 5      & Aug 23 09:20 &   504   &  96.9  &  0.48 & Non-dip \\
 6      & Aug 23 09:36 &   137   &  34.8  &  0.56 & Non-dip \\
 7      & Aug 23 14:55 &  \dots  &  \dots &  0.38 & Non-dip\\
 8      & Aug 23 17:19 &   372   &  84.0  &  0.19 & Dip \\
 9      & Aug 23 19:44 &   292   &  66.5  &  0.01 & Dip \\ 
10      & Aug 23 22:13 &   437   &  82.1  &  0.85 & Non-dip \\
\noalign{\smallskip}
\hline
\end{tabular}
\end{flushleft}
\label{tab:bursts}
\end{table}

In order to investigate the spectral properties of the bursts, 3 intensity 
selected non-dip burst spectra (numbers 3, 4, 5, and 10) 
were first accumulated with count rates of 2--5, 5--8, and $>$8 s$^{-1}$. 
The non-burst, non-dip continuum was again used for background subtraction. 
The 3 spectra were fitted with an absorbed blackbody model and the results 
given in Table~4. The \nh\ values are consistent with 
those obtained for the non-dip continuum, and there is an indication of 
temperature change with burst intensity. A F-test shows that the
temperature changes are real compared with the mean temperature in the
3 spectra at confidence $\>>$99.9\%. Using 
${\rm R_{bb} = d \sqrt ( f_x /(\sigma T^4_{bb}))}$, 
where $\rm {f_x}$ is the flux and d the distance of the source,
to estimate the blackbody
emission radii for each burst spectrum gives similar values of $\sim$3~km,
using the temperatures and luminosities, L, given in 
Table~4. 

\begin{table*}
\caption[]{Intensity selected non-dip burst spectral fit parameters.
90\% confidence limits are given. The luminosity values are bolometric
and assume a distance of 10~kpc}
\begin{flushleft}
\begin{tabular}{llllll}
\noalign{\hrule\smallskip} 
Intensity & Total & N$_{\rm H}$  &   kT${\rm _{b}}$    & Mean L &$\chi^2$/dof \\
(count s$^{-1}$) & count & ($10^{22}$~cm$^{-2}$) & (keV) & (erg ~s$^{-1}$) & \\ 
\noalign{\hrule\smallskip} 
 $>$8   & 965 & $3.0\pm _{1.3}^{1.5}$ & $2.00 \pm 0.22$ & $2.2\times 10^{37}$ & 37/41 \\
 5--8   & 485 & $3.3\pm _{2.6}^{3.1}$ & $1.57\pm 0.22$ & $8.9\times 10^{36}$ & 18/23 \\
 2--5   & 519 & $3.4\pm _{2.2}^{3.0}$ & $1.46\pm 0.24$ & $3.5\times 10^{36}$ & 38/31 \\
 \noalign{\smallskip}
\hline
\end{tabular}
\end{flushleft}
\label{tab:burst_fits}
\end{table*}

Previously, bursts in dips have only been observed during the {\it Ginga}
observations of XB{\thinspace}1916$-$053 when 2 out of 29 bursts occurred
in dips. The spectral properties of these have been studied by Smale et al. (1992)
and by Yoshida (1993) who proposed that the material responsible
for the dipping is temporarily ionized by the bursts. In order
to investigate whether this scenario is consistent with the
burst spectra presented here, a simple spectral comparison between 
burst no. 1 and the sum of the non-dip bursts was performed
using a spectrum extracted from the peak of each burst in each case.  
The first burst was chosen since it 
appears to be most strongly affected of the bursts that 
occurred in dips. No attempt was made to sum data from the
different bursts that occurred in dips, due to differing continuum levels.
and the difficultly in reliably estimating the continuum level to be 
subtracted, due to the high variability during dipping intervals.
The results are shown in 
Fig.~7. In both panels the summed spectrum is well
fitted with an absorbed blackbody.
The left panel also shows the best-fit to burst no. 1 using a 
blackbody with absorption fixed at $3\times 10^{23}$~atom cm$^{-2}$,
a value much less than those we obtained in dip fitting (see Table 2).
As can be seen, such a column density is strongly inconsistent with the burst 
spectrum, providing too much absorption. In constrast, the right panel shows
a fit using a highly ($\xi \approxgt 1000$) ionized absorber which
matches the data well. This suggests that
an ionized absorber is an explanation of the lack of strong absorption
for the bursts in dips.

\begin{figure}
  \begin{flushleft}
  \centerline{\psfig{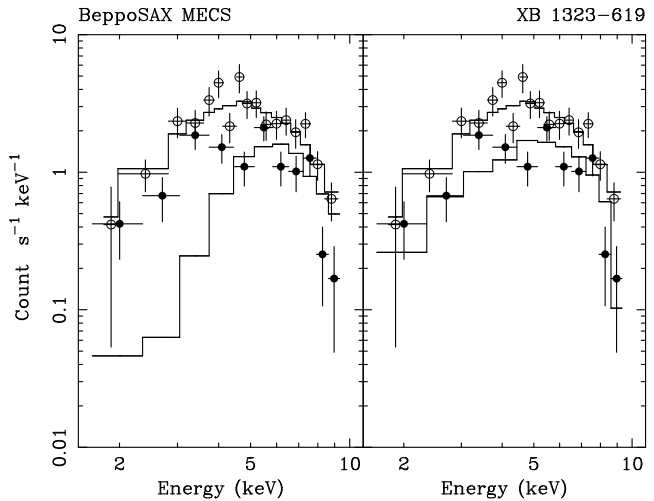}}
  \caption[]{Fits to MECS summed out of dip bursts (open
             circles) and to burst no. 1 that occurred during a dip 
             (filled circles).
             In both panels the out of dip burst spectra are fitted with
             an absorbed blackbody model.   
             The left panel shows a fit to the burst no. 1 spectrum with the 
             absorption of $3 \times 10^{23}$~atom~cm$^{-2}$ 
             and is clearly inconsistent. 
             The right panel shows a fit to the same spectrum
             using an ionized absorber}
  \label{fig:burst_spec}
  \end{flushleft}
\end{figure}

\section{Discussion}

We have observed \src\ in a very broad energy range using \sax. As was found 
in the \sax\ observation of the dipping LMXB XB\thinspace 1916-053 
(Church et al. 1998b), the spectrum of \src\ extends towards the highest 
energies of the PDS. \sax\ has allowed rigorous testing, for the first time,
of absorbed + unabsorbed modelling, and we have shown
that the large change in normalization required in the unabsorbed
component is inconsistent with the relatively small upper limit decrease
of $\rm {10 \pm 10}$\% in the PDS intensity above 20~keV, and so this model cannot
fit the broadband spectral evolution.
However, the non-dip spectrum is well fitted by
the two-component model consisting of point-like blackbody emission
from the surface or boundary layer at the surface of the neutron star, plus
extended Comptonized emission from an Accretion Disk Corona. The dip
spectra can be well fitted by the progressive covering model in which the
point-source blackbody is covered rapidly and the extended emission
is covered progressively as the depth of dipping increases. 
This becomes the third dipping source with an unabsorbed part of the spectrum
that has been modelled in this way (with XB\thinspace 1916-053 and
EXO\thinspace 0748-676; Church et al. 1998a,b).
The systematic increase in covering fraction is consistent with a large
absorber moving across the emission regions. 
The extended nature of the non-thermal Comptonized component is, of
course, strongly indicated by the need for progressive covering of
this component to explain the spectral evolution during dips, and this
is also true in the other sources explained by progressive covering. 
The identification of the extended emission component with emission from the
ADC is supported, for example, by application of this model to
EXO\thinspace0748$-$676, in which it was necessary to add a line at 0.65~keV
to the extended emission component, the line most probably originating in
the ADC (Church et al. 1998a). The column density for the blackbody
increases rapidly implying a point source, and to much higher values than
for the non-thermal emission consistent with the point source being
obscured by higher density regions in the absorber, whereas the extended
emission measures an average across the absorber. The break energy of $\sim$45~keV
is somewhat lower than the value found in XB\thinspace1916$-$053. This
implies an electron temperature in the Comptonizing region of $\sim$17~keV. 

The fits to the dip spectra clearly show that the covering fraction rises 
to only $\sim$ 67\% in deepest dipping in the total MECS band, 
unlike in the sources XB\thinspace1916$-$053 and EXO\thinspace
0748$-$676 in which the covering fraction determined in the same way
often reaches $\sim$100\% in deepest dipping (Church et al. 1997, 1998a,b).

We have investigated whether this difference may be related to 
dust scattering given the non-dip column density in \src\ 
of $\rm {\sim 4\times 10^{22}}$ atom cm$^{-2}$ 
compared with $3.2 \times 10^{21}$~\hcm\ in XB\thinspace 1916-053 (Church et al. 1998b)
and $1.6 \times 10^{21}$~\hcm\ in EXO\thinspace 0748-676 Church et al.
(1998a). 
This high absorption will result in a substantial dust scattering 
halo around \src\ (e.g., Predehl \& Schmitt 1995), unless it is intrinsic 
to the source. For the observed \nh\, the optical extinction, ${\rm A_\nu}$, 
is expected to be $\sim$14~mag. Using the relation for the dust scattering optical depth,
${\rm \tau_{sca}}$, derived by Predehl \& Schmitt of
${\rm \tau_{sca} = 0.087 \times A_\nu \times E_{(keV)}^{-2}}$ gives
${\rm \tau_{sca}= 0.3}$ at 2~keV, or 0.05 at 5~keV. 
The fractional halo intensity, ${\rm I_{frac}}$, is given by
$1 - \exp(-\tau_{sca})$ and is 25\% at 2~keV, or 5\% at 5~keV.
The intensity of any scattered halo will not reflect short-term
changes in the intensity of the central source since such variability
is smoothed by differential time delays of up to days (see e.g.,
Bode et al. 1985). During the $\sim$1~hr duration dipping intervals the scattering
halo will remain visible and its intensity and spectrum will
be determined by the mean spectrum of \src\ over the last days,
as well as the properties of the dust grains themselves. 
However, this assumes that all of the halo is collected requiring
extraction of data from a relatively large circular region 
centred on the source in the imaging instruments compared with the
radii actually used (4\arcmin\ for MECS), so that the amount of halo 
collected may be substantially less than calculated above.
Inspection of Fig.~4 shows that the depth of dipping is $\sim$65\% at 2 keV 
(for the deep dip spectrum), although the maximum depth of dipping in the
light curve will be somewhat greater because of the averaging involved
in accumulating the deep dip spectrum.
However, if we take 35\% as an upper limit to the possible contribution
of dust scattering at 2 keV, the contributions at higher energies will
be reduced by the expected $\rm E^{-2}$ dependence, and the contribution
integrated over the band 2--10~keV can be estimated as $<$ 10\%. Thus,
it is unlikely that dust scattering can explain the fact that the covering
fraction does not exceed 67\% in this band.
A possible explanation of this is that the absorber is ``blobby'', such
that an appreciable fraction of the absorber consists of lower density
regions, 
which effectively do not cover the emission regions reducing the measured 
covering fraction {\it f}. The measured value of {\it f} will be the
product of two terms: $\rm {f_{env} \cdot f_{blob}}$, where $\rm
{f_{env}}$ is the covering fraction between the envelope of the source
and the extended emission region, and $\rm {f_{blob}}$ is the partial
covering fraction in the overlapping, blobby absorber. Alternatively,
the total angular extent of the absorber could be less than that of the
extended emission region. However, if the preliminary result that the
depth of dipping at the lowest energies reaches 100\% is correct,
given the poor statistics of this result, then the absorber has to be
larger than the source. Further work is needed to clarify this point.

The fits to the cutoff power-law and blackbody model indicate that
the blackbody emission has a luminosity of $\rm {2.6\times 10^{35}}$~\ergsec, 
for an assumed distance of 10~kpc. 
However, the blackbody is weak and the uncertainties in $\rm {kT_{bb}}$
may be larger than shown in Table 1.
The blackbody comprises 14\% of the 
2--10~keV flux and the blackbody radius $\rm {R_{bb}}$ is 0.45 km
suggesting that this emission originates in a small boundary layer on the
neutron star where the accretion disk impacts. Rutledge et al. (1999)
however, have shown that a simple blackbody model may underestimate the
emitting area. The total
1--10 keV luminosity is $\rm {1.9\times 10^{36}}$ \ergsec. These
values may be compared with recent results from \sax\ for 
XB\thinspace 1916$-$053 in which the blackbody luminosity was 
$\rm {5 \times 10^{35}}$ \ergsec, the blackbody 20\% of the 1--10~keV flux, the
blackbody radius 0.76 km and the total 1--10 keV luminosity 
$\rm {3\times 10^{36}}$ \ergsec (Church et al. 1998b). It can be seen that the
luminosities are scaled down in \src\ by about a factor of 2 compared
with XB\thinspace 1916$-$053, with an equivalent change in blackbody
radius. 
The diameter of the extended emission component 
$\rm {d_{ADC}}$ can be estimated from the dip ingress and egress
timescales $\sim $250--500~s. 
Assuming that the absorber is more extended than the source region,
$\rm {d_{ADC}}$ is given by
$\rm {2\,\pi\,r_{disk}\, \Delta t/P}$ where $\rm {\Delta t}$ is the transition
time, r$_{disk}$ is the radius of the accretion disk
and P is the orbital period. For $\rm {\Delta t}$ equal to 250--500~s, 
$\rm {d_{ADC}}$ is 4--8$\rm {\times 10^9}$ cm, assuming a disk radius of
$\rm {3\times 10^{10}}$~cm. 

The bursts occurring in dips provide burst and absorber diagnostics that are
not normally available. Previously, 2 bursts in dips have been observed
in the {\it Ginga} observations of XB\thinspace 1916$-$053
(Smale et al. 1992; Yoshida 1993). The bursts were 
not heavily  absorbed, which Smale et al. and Yoshida attribute 
to the almost instantaneous ionization 
of the material responsible for the dipping. This material then
returned to its equilibrium state on the same time scale as the
burst decay. Smale et al. (1992) demonstrate that the burst fluence
is sufficient to cause significant photoionization at the outer radius 
of the disk of $1 \times 10^{10}$~cm in the case of
XB{\thinspace}1916$-$053. 
In \src\, the calculated radius of the accretion disk $\rm {r_{disk}}$ is 
$\rm {\sim 3 \pm 1\times 10^{10}}$ cm. A value of the ionization parameter, 
$\xi = {\rm L/nr^2} \approxgt 500$ is required for absorption to be
markedly reduced compared with a neutral absorber, where r is the 
distance of the obscuring material from the central X-ray source and
n its density. If the absorbing region has a thickness along the line of
sight of $\rm {\epsilon \cdot r_{disk}}$, then 
$\xi = {\rm L\, \epsilon/ N_H\, r_{disk}}$ where \nh\ is the column 
density due to the absorber, not including interstellar absorption. 
The value of $\epsilon$ is difficult to assess, 
but values of 0.1 have been suggested (e.g. Mason et al. 1985).
However, in the present case, 
dipping lasts 40\% of the orbital cycle implying a very extended absorber, 
and it is possible that
$\epsilon$ could be as much as 0.3.
The mean burst luminosity close to the peaks of the bursts is 
$2.7 \times 10^{37}$~\ergsec. 
Spectral fitting gives values for \nh\ as an average across the absorbing
region of $\rm {\sim 20\times 10^{22}}$ atom cm$^{-2}$ after subtraction of
the interstellar contribution (see Table 2). Thus the mean value of $\xi$, equivalent
to the mean column density for the absorber, is $\sim$500
for $\epsilon$ = 0.1, and 1500 for $\epsilon$ = 0.3. It thus appears
likely, given the large uncertainties in this calculation, in L, 
$\rm {r_{disk}}$
and $\epsilon$, that the absorber is substantially ionized during bursts 
so that bursts in dips will not show a strong increase in absorption.

In this case, the reduction of integrated count of $\sim$30\% of the burst in
the dip (compared with non-dip bursts) will be due to electron
scattering. For the first burst which coincides with a dip, the
intensity is 9 count s$^{-1}$ compared with a height of 17 count
s$^{-1}$ for the non-dip bursts. This reduction requires an electron
column density $\rm {N_e}$ of $\rm {\sim 100\times 10^{22}}$ electron
$\rm {cm^{-2}}$. This value can be compared with values of the column
density for the blackbody in dipping, i.e. the column density
applicable to the point-source neutron star, which should provide the
electron column density applicable to bursts on the neutron star. However
out of bursts, using the maximum $\rm
{N_H}$ of the extended emission component as the average of
the absorber ($\rm {\sim 25\times 10^{22}}$ atom cm$^{-2}$; Table 2),
the average $\xi $ of the absorber is $\sim$25 showing that hydrogen is
ionized. The reduction in intensity that this would cause due to
electron scattering is $\sim$15\%, which would be difficult to detect
as a vertical shift in PDS data (between non-dip and deep dip) 
but is consistent with our upper limit of $\rm {10 \pm 10}$\%
between 20 and 50 keV. The consequence of this is
that during a burst, the other elements may become fully ionized, but
the electron density $\rm {n_e}$ would increase less than 20\%
assuming Solar abundances. The burst
referred to occupied the second half of the dip containing it, so 
the appropriate column density will be less than the maximum blackbody
value, and we estimate $\rm {500 - 1000\times 10^{22}}$ atom cm$^{-2}$
may be appropriate. Thus the extra $\rm {N_e}$ may be $\rm {\sim
150\times 10^{22}}$ electrons cm$^{-2}$. This compares well with the
value required to reduce the burst intensity as observed. 

\begin{acknowledgements}
The \sax\ satellite is a joint Italian-Dutch programme.
We thank the staff of the \sax\ Science Data Center
for help with these observations. M. Ba\l uci\'nska-Church and M.J. Church
thank the Astrophysics Division of ESA for their hospitality during a
recent visit to ESTEC to work on this data.
\end{acknowledgements}


\begin{thebibliography}{}
\bibitem[1989]{}
Anders E., Grevesse N., 1989, Geochimica et Cosmochimica Acta 53, 197

\bibitem[1997]{}
Angelini L., Church M.J., Ba\l uci\'nska-Church M., Parmar A.N., 
Mineo T., 1998, A\&A 339, L41

\bibitem[1997]{}
Bode M.F., Priedhorsky W.C., Norwell G.A., Evans A. 1985, ApJ 299, 845

\bibitem[1997]{}
Boella G., Chiappetti L., Conti G., et al., 1997, A\&AS 122, 327

\bibitem[1993]{}
Church M.J., Ba\l uci\'nska-Church M., 1993, MNRAS 260, 59

\bibitem[1995]{}
Church M.J., Ba\l uci\'nska-Church M., 1995, A\&A 300, 441

\bibitem[1997]{}
Church M.J., Dotani T., Ba\l uci\'nska-Church M., et al., 1997, ApJ 491, 388

\bibitem[1998]{}
Church M.J., Ba\l uci\'nska-Church M., Dotani T., Asai K., 1998a, ApJ
504, 516

\bibitem[1998]{}
Church M.J., Parmar A.N., Ba\l uci\'nska-Church M., et al., 1998b, A\&A,
338, 556

\bibitem[1986]{}
Courvoisier T.J.-L., Parmar A.N., Peacock A., Pakull M., 1986,
ApJ 309, 265

\bibitem[1976]{}
Forman W., Jones C., Cominsky L., et al., 1978, ApJS 38, 357

\bibitem[1976]{}
Frontera F., Costa E., Dal Fiume D., et al., 1997, A\&AS 122, 371

\bibitem[1999]{}
Jonker P.G., van der Klis M., Wijnands R., 1999, ApJL, in press

\bibitem[1976]{}
Manzo G., Guarrusso S., Santangelo A., et al., 1997, A\&AS 122, 341

\bibitem[1976]{}
Mason K.O., Parmar A.N., White N.E., 1985, MNRAS 216, 1033

\bibitem[1976]{}
Morrison D., McCammon D., 1983, ApJ 270, 119

\bibitem[1976]{}
Parmar A.N., White N.E. 1988,  In: Pallavacini R., White N.E.
(eds.) X-ray Astronomy with EXOSAT. Journal of the Italian 
Astronomical Society 59, p.~147

\bibitem[1986]{}
Parmar A.N., White N.E., Giommi P., Gottwald M., 1986,
ApJ 308, 199

\bibitem[1989]{}
Parmar A.N., Gottwald M., van der Klis M., van Paradijs J., 1989, ApJ
338, 1024 

\bibitem[1997]{} 
Parmar A.N., Martin D.D.E., Bavdaz M., et al., 1997, A\&AS 122, 309

\bibitem[1995]{}
Predehl P., Schmitt J.H.M.M., 1995, A\&A 293, 889

\bibitem[]{}
Rutledge R.E., Bildsten L., Brown E.F., Pavlov G.G., Zavlin V.E.,
1999, ApJ, in press

\bibitem[1995]{}
Smale A.P., 1995, AJ 110, 1292

\bibitem[1992]{}
Smale A.P., Mukai K., Williams O.R., Jones M.H.
Corbet R.H.D., 1992, ApJ 400, 330

\bibitem[1992]{}
Stark A.A., Gammie C.F., Wilson R.W., et al., 1995, ApJS 79, 77

\bibitem[1994]{}
Tanaka Y., Inoue H., Holt S.S., 1994, PASP 46, L37

\bibitem[1992]{}
Van der Klis M., Jansen F., van Paradijs J., Stollman G., 1985, 
Space Sci. Rev. 30, 512

\bibitem[1982]{}
Warwick R.S., Marshall N., Fraser G.W., et al., 1981, MNRAS 197, 865

\bibitem[1982]{}
White N.E., Swank J.H., 1982, ApJ 253, L61

\bibitem[1993]{}
Yoshida K., 1993, {\it Ph.D. thesis} Tokyo University.


\end{thebibliography}
\end{document}